\documentclass[preprint,showpacs,preprintnumbers,amsmath,amssymb]{revtex4}

\usepackage[dvips]{graphicx}
\usepackage{subfigure}

\begin{document}

\author{L. M. Le\'on Hilario}
\author{A. Bruchhausen}
\author{A. M. Lobos}
\author{A. A. Aligia}
\title{Theory of polariton mediated Raman scattering in microcavities}


\date{\today}

\begin{abstract}
We calculate the intensity of the polariton mediated inelastic light
scattering in semiconductor microcavities. We treat the exciton-photon
coupling nonperturbatively and incorporate lifetime effects in both excitons
and photons, and a coupling of the photons to the electron-hole continuum.
Taking the matrix elements as fitting parameters, the results are
in excellent agreement with measured Raman intensities due to optical phonons
resonant with the upper polariton branches in II-VI microcavities with
embedded CdTe quantum wells.

\end{abstract}

\affiliation{Centro At\'omico Bariloche and Instituto Balseiro, Comisi\'on Nacional de
Energ\'{\i}a At\'omica, 8400 S. C. de Bariloche, Argentina}

\maketitle


Planar semiconductors microcavities (MC's) have attracted much attention
in the last decade as they provide a novel means
to study, enhance and control the interaction between light and matter
\cite{Libro-CavityPolaritons(03),
SemicondSciTech18-issue10(03),sijpcm, Skolnick_SST98,
Kasprzak_Nature06, Deng-PRL97-146401(06), Trigo}.
When the MC mode (cavity-photon) is tuned in near resonance with the embedded
quantum-well (QW) exciton transitions, and the damping processes involved
are weak in comparison to the photon-matter interaction, the eigenstates of
the system become mixed exciton-photon states, \textit{cavity-polaritons},
which are in part light and in part matter bosonic quasi-particles
\cite{Libro-CavityPolaritons(03), SemicondSciTech18-issue10(03), sijpcm}.
Examples of interesting new physics are the recent evidence of a
Bose-Einstein condensation of polaritons in CdTe MC's
\cite{Kasprzak_Nature06, Deng-PRL97-146401(06)}, and the construction of devices which increase
the interaction of sound and light, opening the possibility of realizing
a coherent monochromatic source of acoustic phonons \cite{Trigo}.

\begin{figure}[htbp]
    \includegraphics[width=1.0\linewidth]{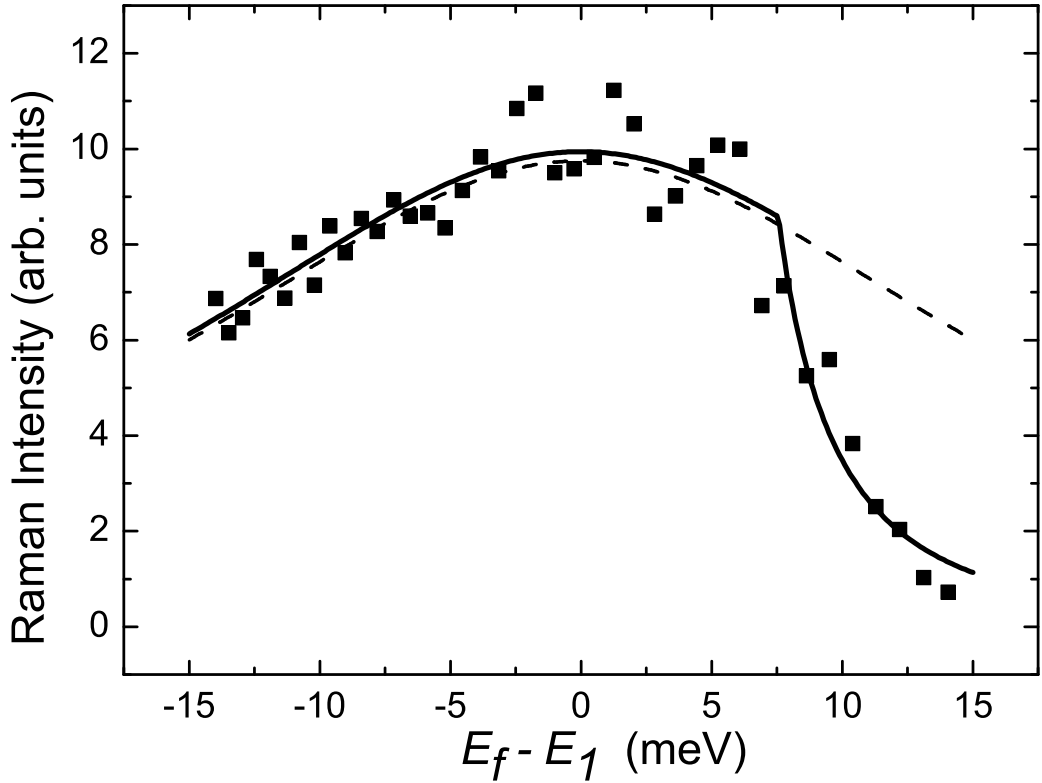}
    \caption{Raman intensity as a function of detuning
for sample A. Solid squares: experimental results
\cite{Bruchhausen_PRB03}. Solid line: theory (Eq. \ref{i5}) for
$B-E_1=14$ meV,
$\delta_{f}= \delta_{1}=0.1$ meV, and $A=0.031$. Dashed
line: result for a 2x2 matrix (Eq. \ref{i2}).}
    \label{rrs1}
\end{figure}

Raman scattering due to longitudinal optical (LO) phonons, being a
coherent process is intrinsically connected with the
cavity-polariton. The physics of strongly coupled photons and
excitons, the polariton--phonon interaction, and the
polariton--external-photon coupling are clearly displayed
\cite{Fainstein_PRL97, Fainstein_PRB98, Bruchhausen_PRB03,
Tribe_PRB97, Stevenson_PRB03}. In particular resonant Raman
scattering (RRS) experiments, in which the wave-length of the
incoming radiation is tuned in a way such that, after the emission
of a LO phonon, the energy of the outgoing radiation coincides
with that of the cavity polariton branches, have proven to be
suited to sense the dynamics of the coupled modes, and to obtain
information about the dephasing of the resonant polaritonic state
\cite{Fainstein_PRL97, Tribe_PRB97, Fainstein_PRB98,
Stevenson_PRB03, Bruchhausen_PRB03}. Unfortunately, due to the
large remaining luminesence, RRS experiments in resonance with the
lower polariton branch have not yet been achieved in intrinsic
II-VI MC's \cite{note1}. Therefore all reported experiments in
these kind of MC's, with embedded CdTe QW's, consider the case of
the scattered photons in outgoing resonance with the upper
polariton branch (for two branch-systems: coupling of one exciton
mode and the cavity photon) or with the middle polariton branch
(for tree branch-systems: coupling of two exciton modes and the
cavity photon) \cite{Fainstein_PRB98,Bruchhausen_PRB03,bru6}.
The measured intensities in these two systems were analyzed on the basis
of a model in which one (two) exciton states $|e\rangle $ are mixed with a
photon state $|f\rangle $ with the same in-plane wave vector $\textbf{k}$,
leading to a 2x2 (3x3) matrix. Essentially, the Raman intensity is proportional to
\begin{equation}
I\sim T_{i}T_{s}|\langle P_{i}|H^{\prime }|P_{s}\rangle |^{2},  \label{i1}
\end{equation}
where $T_{i}$ describe the probability of conversion of an incident photon
$|f_{i}\rangle $ into the polariton state $|P_{i}\rangle $, $T_{s}$ has an
analogous meaning for the scattered polariton $|P_{s}\rangle $ and the
outgoing photon $|f_{s}\rangle $, and $H^{\prime }$ is the interaction
between electrons and the LO phonons. At the conditions of resonance with
the outgoing polariton, $T_{i}$ is very weakly dependent on laser energy or
detuning (difference between photon and exciton energies), while $T_{s}$ is
proportional to the photon strength of the scattered polariton $|\langle
f_{s}|P_{s}\rangle |^{2}$. Similarly, in the simplest case of only one
exciton (2x2 matrix) one expects that the matrix element entering
Eq. (\ref{i1}) is proportional to the exciton part of the scattered polariton
$|\langle P_{i}|H^{\prime }|P_{s}\rangle |^{2}\sim |\langle
e_{s}|P_{s}\rangle |^{2}$. Thus if the wave function is $|P_{s}\rangle =\alpha
|f_{s}\rangle +\beta |e_{s}\rangle $, one has
\begin{equation}
I\sim |\langle f_{s}|P_{s}\rangle |^{2}|\langle e_{s}|P_{s}\rangle
|^{2}=|\alpha |^{2}|\beta |^{2}.  \label{i2}
\end{equation}
This model predicts a Raman intensity which is symmetric with
detuning and is maximum at zero detuning. In other words, the
intensity is maximum for detunings such that the scattered
polariton is more easily coupled to the external photons, but at
the same time when the polariton is more easily coupled to the
optical phonons, which requires a large matter (exciton) component
of the polariton. This result is in qualitative agreement with
experiment \cite{Bruchhausen_PRB03}. However, for positive
detuning the experimental results fall bellow the values predicted
by Eq. (\ref{i2}) (see Fig.~\ref{rrs1}). This is ascribed to the
effects of the electron-hole continuum above the exciton energy,
which are not included in the model \cite{Bruchhausen_PRB03}.

For another sample in which two excitons are involved, the above analysis
can be extended straightforwardly and the intensity depends on the
amplitudes of a 3x3 matrix and matrix element of $H^{\prime }$ involving
both excitons \cite{Bruchhausen_PRB03}. However, comparison with experiments
at resonance with the middle polariton branch, shows a poorer agreement than
in the previous case (Fig. 6 of Ref. \onlinecite{Bruchhausen_PRB03}). In addition,
a loss of coherence between the scattering of both excitons with the LO phonon
was assumed, which is hard to justify. Some improvement has been obtained
recently when damping effects are introduced phenomenologically as
imaginary parts of the photon and exciton energies,
but still a complete loss
of coherence resulted from the fit \cite{Bruchhausen_AIP05,bru6}.

In this paper we include the states of the electron-hole continuum, and the
damping effects in a more rigorous way. Using some matrix elements as free
parameters, we can describe accurately the Raman intensities for both
samples studied in Ref. \onlinecite{Bruchhausen_PRB03}.

In the experiments with CdTe QW's inside II-VI MC's, the light
incides perpendicular to the $(x,y)$ plane of the QW's and is
collected in the same direction $z$. Therefore the in-plane wave
vector $\mathbf{K}=0$, the polarization of the electric field
should lie in the $(x,y)$ plane, and the excitons which couple
with the light should have the same symmetry as the electric field
(one of the two $\Gamma _{5}$ states of heavy hole excitons
\cite{Jorda_PRB93}). Thus, to lighten the notation we suppress
wave vector and polarization indices. The basic ingredients of the
theory are two or three strongly coupled boson modes, one for the
light MC eigenmode with
boson creation operator $f^{\dagger }$, another one for the $1s$ exciton ($%
e_{1}$) and if is necessary, the $2s$ exciton ($e_{2}$) is also
included. We assume that each of these boson states mixes with a
continuum of bosonic excitations which broadens its spectral
density. In addition, we include the electron-hole continuum above
the exciton states, described by bosonic operators
$c_{\textbf{k}}^{\dagger },c_{\textbf{k}}$, where $\textbf{k}$ is
the difference between electron and hole momentum in the $(x,y)$
plane (the sum is $\mathbf{K}=0$ because it is conserved).

The Hamiltonian reads:
\begin{eqnarray}
H &=&E_{f}f^{\dagger }f+\sum_{i}E_{i}e_{i}^{\dagger
}e_{i}+\sum_{i}(V_{i}e_{i}^{\dagger }f+\text{H.c.})
+\sum_{p}\epsilon _{p}r_{p}^{\dagger }r_{p}+\sum_{p}(V_{p}r_{p}^{\dagger
}f+\text{H.c.})  \nonumber \\
&+&\sum_{i\textbf{q}}\epsilon
_{i\textbf{q}}d_{i\textbf{q}}^{\dagger
}d_{i\textbf{q}}+\sum_{i\textbf{q}}(V_{i\textbf{q}}d_{i\textbf{q}}^{\dagger
}e_{i}+\text{H.c.}) +\sum_{\textbf{k}}\epsilon
_{\textbf{k}}c_{\textbf{k}}^{\dagger
}c_{\textbf{k}}+\sum_{\textbf{k}}(V_{\textbf{k}}c_{\textbf{k}}^{\dagger
}f+\text{H.c.}). \label{mod}
\end{eqnarray}
The first three terms describe the strong coupling between the MC photon and
the exciton(s) already included in previous approaches \cite%
{Bruchhausen_PRB03,Bruchhausen_AIP05}. The fourth and fifth terms
describe a continuum of radiative modes and its coupling to the MC
light eigenmode. Their main effect is to broaden the spectral
density of the latter even in the absence of light-matter
interaction. The following two terms have a similar effect for the
exciton mode(s). The detailed structure of the states described by
the $d_{i\textbf{q}}^{\dagger }$ operators is not important in
what follows. They might describe combined excitations due to
scattering with acoustical phonons. The last two terms correspond
to the energy of the electron hole excitations and their coupling
to the MC light mode.

In Eq. (\ref{mod}) we are making the usual approximation of neglecting the
internal fermionic structure of the excitons and electron-hole operators and
taking them as free bosons. This is an excellent approximation for the
conditions of the experiment. We also neglect terms which do not conserve the
number of bosons. Their effect is small for the energies of interest \cite%
{Jorda_PRB94}. These approximations allow us formally to
diagonalize the Hamiltonian by a Bogoliubov transformation. The
diagonalized Hamiltonian has the form $H=\sum_{\nu }E_{\nu }p_{\nu
}^{\dagger }p_{\nu }$, where the boson operators $p_{\nu
}^{\dagger }$ correspond to generalized polariton operators and
are linear combinations of all creation operators entering Eq.
(\ref{mod}). Denoting the latter for brevity as $b_{j}^{\dagger
}$, then $p_{\nu }^{\dagger }=\sum_{j}A_{\nu j}b_{j}^{\dagger }$.
In practice, instead of calculating the $E_{\nu }$ and $A_{\nu j
}$, it is more convenient to work with retarded Green's functions
$G_{jl}(\omega )=\langle \langle b_{j};b_{l}^{\dagger }\rangle
\rangle _{\omega }$ and their equations of motion
\begin{equation}
\omega \langle \langle b_{j};b_{l}^{\dagger }\rangle \rangle _{\omega
}=\delta _{jl}+\langle \langle [b_{j},H];b_{l}^{\dagger }\rangle \rangle
_{\omega }.  \label{em}
\end{equation}
As we show below, the RRS intensity can be expressed in terms of spectral
densities derived from these Green's functions, which in turn can be
calculated using Eq. (\ref{em}).

Using Fermi's golden rule, the probability per unit time for a transition
from a polariton state $|i\rangle =p_{\nu ^{\prime }}^{\dagger }|0\rangle $
to states $|s\rangle =p_{\nu }^{\dagger }a^{\dagger }|0\rangle $, where $%
a^{\dagger }$ creates a LO phonon is
\begin{equation}
W=\frac{2 \pi}{\hbar}|\langle i|H^{\prime }|s\rangle |^{2}\rho (\omega ),
\;\;\text{where}\;\;
\rho (\omega )=\sum_{\nu }\delta (\omega -E_{\nu })=-\frac{1}{\pi }%
\text{Im}G_{\nu \nu }(\omega +i0^{+})
\end{equation}
is the density of final states. As argued above, we
neglect the dependence of $T_{i}$ on frequency and take $T_{s}=|A_{\nu
f}|^{2}$, the weight of photons in the scattered eigenstates. In addition,
$H^{\prime }$ should be proportional to the matter
(exciton) part of the scattered polariton states. Then, the Raman intensity
is proportional to:
\begin{equation}
WT_{s} \propto I=|A_{\nu e1}+\alpha A_{\nu e2}|^{2}|A_{\nu f}|^{2}
\rho (\omega ).  \label{i3}
\end{equation}
Here we are neglecting the contribution of the electron-hole continuum to $%
H^{\prime }$, and $\alpha $ is the ratio of matrix elements of the
exciton-LO phonon interaction between $2s$ and $1s$ excitons. If the
$2s$ excitons are unimportant, $\alpha =0$.

Using Eqs. (\ref{em}) it can be shown that
\begin{eqnarray}
\rho _{jl}(\omega ) &=&-\frac{1}{2\pi }[G_{jl}(\omega +i0^{+})-G_{jl}(\omega
-i0^{-})] =A_{\nu j}\bar{A}_{\nu l}\rho (\omega ).  \label{roc}
\end{eqnarray}
From here and $\sum_{j}|A_{\nu j}|^{2}=1$, it follows that
$\rho (\omega )=\sum_{j}\rho _{jj}(\omega )$. Replacing in Eq. (\ref{i3}) we
obtain:
\begin{equation}
I(\omega )=\frac{\rho _{ff}[\rho _{e1,e1}+|\alpha |^{2}\rho _{e2,e2}+2\text{%
Re}(\alpha \rho _{e2,e1})]}{\sum_{j}\rho _{jj}(\omega )}.  \label{i4}
\end{equation}
In practice, when $\omega $ is chosen such that the resonance condition for
the outgoing polariton is fulfilled, we can neglect the contribution of the
continuum states in the denominator of Eq. (\ref{i4}). In particular, if the
contribution of the $2s$ exciton can be neglected (as in sample A of Ref. %
\onlinecite{Bruchhausen_PRB03})
\begin{equation}
I(\omega )=\frac{\rho _{ff}(\omega )\rho _{e1,e1}(\omega )}{\rho
_{ff}(\omega )+\rho _{e1,e1}(\omega )}.  \label{i5}
\end{equation}
The Green's functions are calculated from the equations of motion (\ref{em}).
In the final expressions, the continuum states enter through the following
sums:
\begin{eqnarray}
S_{f}(\omega ) &=&\sum_{p}\frac{|V_{p}|^{2}}{\omega
+i0^{+}-\epsilon _{p}}, \;\; S_{i}(\omega
)=\sum_{\textbf{q}}\frac{|V_{i\textbf{q}}|^{2}}{\omega +i0^{+}
-\epsilon_{i\textbf{q}}}, \;\;S_{f}^{\prime }(\omega )
=\sum_{\textbf{k}}\frac{|V_{\textbf{k}}|^{2}}{\omega
+i0^{+}-\epsilon _{\textbf{k}}} \label{sums}
\end{eqnarray}
For the first two we assume that the results are imaginary constants that we
take as parameters:
\begin{equation}
S_{f}(\omega )=-i\delta _{f},\;\;\;\;\ S_{j}(\omega )=-i\delta _{j}
\label{deltas}
\end{equation}
This is the result expected for constant density of states and
matrix elements. Our results seem to indicate that this assumption
is valid for the upper and middle polariton branches. For the
lower branch at small $\textbf{k}$ it has been shown that the line
width due to the interaction of polaritons with acoustic phonons
depends on detuning $E_{f}-E_{1}$ and $k_{\parallel }$, being
smaller for small wave vector \cite{Savona_PSS97,Cassabois_PRB00}.

The electron-hole continuum begins at the energy of the gap and
corresponds to vertical transitions in which the light promotes a
valence electron with 2D wave vector $\textbf{k}$ to the
conduction band with the same wave vector. In the effective-mass
approximation, the energy is quadratic with $\textbf{k}$ and this
leads to a constant density of states beginning at the gap. The
matrix element $V_{\textbf{k}}$ is proportional to
$M_{\textbf{k}}=\langle \textbf{k}_{v}|p_{E}|\textbf{k}_{c}\rangle
$, where $p_{E}$ is the momentum operator in the direction of the
electric field, and $|\textbf{k}_{v}\rangle $,
$|\textbf{k}_{c}\rangle $ are the wave functions for valence and
conduction electrons with wave vector $\textbf{k}$. Taking these
wave functions as plane waves, one has $M_{\textbf{k}}\sim k_{E}$,
the wave vector in the direction of the electric field. Then
$|V_{\textbf{k}}|^{2}\sim k_{E}^{2}$. Adding the contributions of
all directions of $\textbf{k}$ one has $|V_{\textbf{k}}|^{2}\sim
|\textbf{k}|^{2}\sim \epsilon _{\textbf{k}}$ (linear with energy
for small energy). This leads to
\begin{equation}
S_{f}^{\prime }=R(\omega )-iA(\omega -B) \Theta (\omega -B),
\label{sumac}
\end{equation}
where $B$ is the bottom of the electron-hole continuum (the energy
of the semiconductor gap) and $A$ is a dimensionless parameter
that controls the magnitude of the interaction. The real part
$R(\omega )$ can be absorbed in a renormalization of the photon
energy and is unimportant in what follows. The imaginary part is a
correction to the photon width for energies above the bottom of
the continuum.


Using the theory outlined above, we calculated the intensity of RRS
corresponding to the samples A and B measured in Refs.
\onlinecite{Bruchhausen_PRB03,bru6}
and compared them with the experimental results.

Sample A corresponds to the simplest case. Two polariton branches
are seen and therefore only the $1s$ exciton plays a significant
role. The binding energy of this exciton $B-E_{1}$ is not well
known. The Rabi splitting $2V_{1}=19$ meV. The width of the Raman
scan as a function of frequency for zero detuning is of the order
of $w=0.1$ meV (see Fig. 3 of Ref.
\onlinecite{Bruchhausen_PRB03}). This implies the relation
$2w^{2}=\delta _{f}^{2}+\delta _{1}^{2}$ in our theory. In any
case the results are weakly sensitive to $w$. Therefore, we have
three free parameters in our theory in addition to a
multiplicative constant: $B-E_{1}$, the ratio of widths
$\delta_{f}/\delta_{1}$ and the slope $A$.

The comparison between the experimental and the theoretical intensities is
shown in Fig.~\ref{rrs1} for a set of parameters that lead to a close agreement
with experiment. The condition of resonance is established choosing the
energy $\omega $ for which the intensity given by Eq. (\ref{i5}) has its
second relative maximum (corresponding to the upper polariton branch). The
dashed line corresponds to the case in which only the first three terms
(with $i=1$) in the Hamiltonian, Eq. (\ref{mod}) are included. In this case,
the intensity is given in terms of the solution of a 2x2 matrix
[Eq. (\ref {i2})] and was used in Ref.  \onlinecite{Bruchhausen_PRB03}
to interpret
the data. This simple expression gives a Raman intensity which is an even
function of detuning. When the full model
is considered, the Raman intensity falls more rapidly for large
detuning $E_{f}-E_{1}$ as a consequence of the hybridization of
the photon with the electron-hole continuum.
When the energy of the polariton increases beyond $B$
entering the electron-hole continuum (corresponding to the kink
in Fig. 1), the Raman peak broadens and loses intensity. The kink can be
smoothed if the effect of the infinite excitonic levels below the continuum
is included in the model (leading to an $S_{f}^{\prime }$ with continuous first
derivative), but this is beyond the scope of this work.
If the ratio
$\delta_{f}/\delta _{1}$ is enlarged, the  Raman intensity increases for negative
detuning with respect to its value for positive detuning.

\begin{figure}[htbp]
    \includegraphics[width=1.0\linewidth]{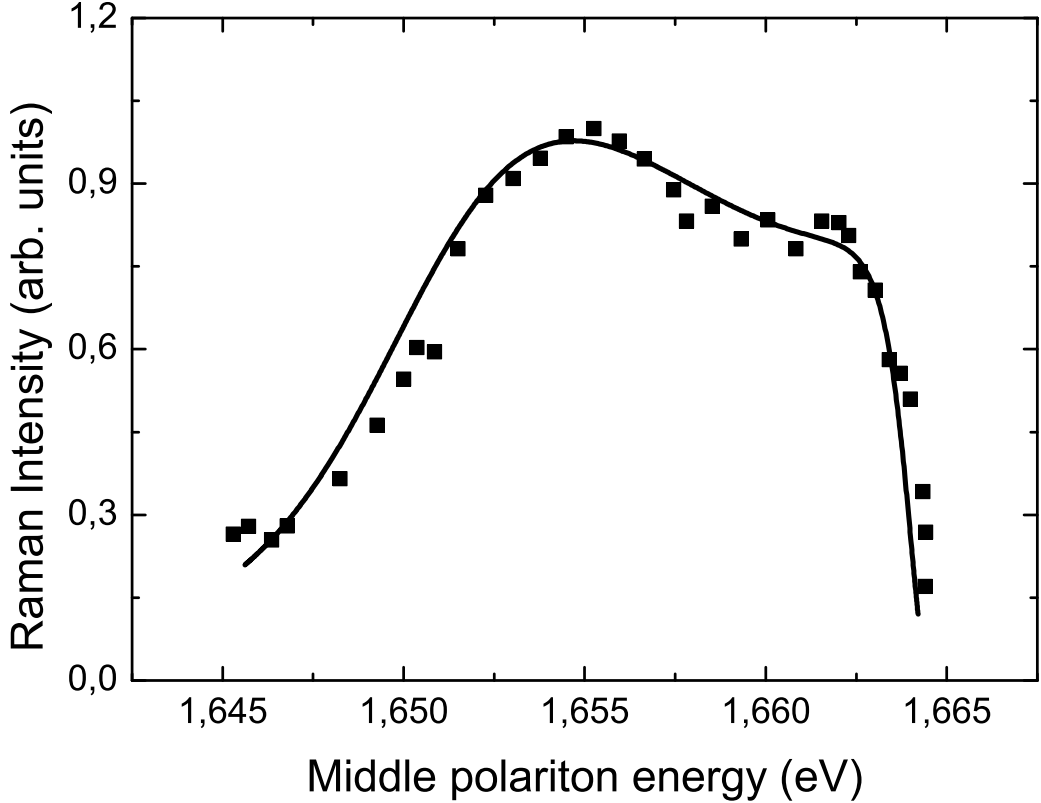}
    \caption{Raman intensity as a function of the middle polariton
energy. Solid squares: experimental results \cite{bru6}. Solid line: theory for
$\delta_{f}=0.2$ meV, $\delta_{1}=0.1$ meV, $\delta_{2}=0.12$ meV,
$\alpha=-0.45$ and $z=0.14$. $A$ is the same as in Fig.~\ref{rrs1} .}
    \label{rrs2}
\end{figure}

In the experiments with sample B three
polariton branches are observed \cite{bru6} and the $2s$ exciton plays a role.
Experimentally, it is known that the binding energy for the two
excitons are $B-E_{1}=$17 meV and $B-E_{2}=$2 meV.
From the observed
Rabi splitting one has $2V_{1}=13$ meV, $2V_{2}=2.5$ meV. In
comparison with the previous case, we have the additional parameter
$\alpha$ (the ratio of exciton-LO phonon matrix elements). In
addition, to be able to describe well the intensity for low
energies of the middle polariton (left part of the curve shown in
Fig.~\ref{rrs2}), we need to assume a small linear dependence of
$E_{1}$ with the position of the incident laser spot in the
sample. This dependence is also inferred form the observed
luminescence spectrum \cite{bru6}.
In our model, this corresponds to a dependence of $E_{1}$ with $%
E_{f}$:
\begin{equation}
E_{1}=E_{1}^{0}+z(E_{f}-E_{1}^{0})  \label{e1vsef}
\end{equation}

In Fig.~\ref{rrs2} we show the intensity at the second maximum of $I(\omega )$
(corresponding to the middle polariton branch) as a function of the energy
of this maximum. We also show in the figure experimental results taken at
lower laser excitation and a slightly higher temperature (4.5 K) than
those reported in Ref. \onlinecite{Bruchhausen_PRB03}.

The slope which better describes the data is $z=0.14$. This value is close to
$z=0.155$ which was obtained from a fit of the maxima of luminescence
spectrum of the lower and middle polaritons. We have taken the same value for $A$ as
in Fig.~\ref{rrs1}. The agreement between theory and experiment is remarkable.
As for sample A, the values of $\delta _{i}$ that result from the fit are
reasonable in comparison with calculated values \cite{Savona_PSS97}.

In summary, we have proposed a theory to calculate Raman intensity for excitation
of longitudinal optical phonons in microcavities, in which different matrix
elements are incorporated as parameters of the model. The most important
advance in comparison with previous simplified theories \cite%
{Bruchhausen_PRB03,Bruchhausen_AIP05} is the inclusion of the strong
coupling of the electron-hole continuum with the microcavity photon.
Inclusion of this coupling is essential when the energy of the polariton is
near the bottom of the conduction band (at the right of Fig. 1). We also
have included the effects of damping of excitons and photons, coupling them
with a continuum of bosonic excitations. Simpler approaches have included
the spectral widths $\delta _{f}$ and $\delta _{i}$ of photons and excitons
as imaginary parts of the respective energies, leading to non-hermitian
matrices.

Taking some of the parameters of the model as free
($\delta _{f}/ \delta _{i}$, $B-E_1$ and $A$ for sample A,  $\delta _{f}$,
$\delta _{i}$, $z$ and $\alpha$ for sample B), we obtain excellent fits of the
observed Raman intensities. The resulting values of the parameters
agree with previous estimates, if they are available. We are not aware
of previous estimates for $A$ and $\alpha$.
As an important improvement to previous approaches
\cite{Bruchhausen_PRB03,Bruchhausen_AIP05} for the case of sample B,
we do not have to assume a partial loss of coherence between
$1s$ and $2s$ excitons in their scattering with the LO phonon.

Further progress in the understanding of the interaction of excitons with
light and phonons requires microscopic calculations of the parameters
$\delta_{f},\delta _{i},A$ and $\alpha $,
and the effects of the temperature on
them. However, taking into account the difficulties in calculating these
parameters accurately, the present results are encouraging and suggest that
the main physical ingredients are included in our model.


We thank A. Fainstein for useful discussions. This work was supported by
PIP 5254 of CONICET and PICT 03-13829 of ANPCyT.


\begin{thebibliography}{99}
\bibitem{note1} RRS experiments in resonance with the lower polariton branch have only been reported in III-VI samples with doped Bragg reflectors (see refs. \cite{Tribe_PRB97} and \cite{Stevenson_PRB03} for details).
\end{thebibliography}
\end{document}